\documentclass[twocolumn,showpacs,preprintnumbers,superscriptaddress,a4paper]{revtex4}

\usepackage{amsmath}
\usepackage{amssymb}
\usepackage{subfigure}
\usepackage{epsfig}

\begin{document}

\preprint{APS/123-QED}

\title{
Electronic and magnetic properties of substitutional Mn clusters in (Ga,Mn)As
}
\author{Hannes Raebiger}
\affiliation{
COMP/Laboratory of Physics, 
Helsinki University of Technology, POB 1100, 02015 HUT, Finland.
}
\author{Andr\'es Ayuela}
\affiliation{Donostia International Physics Centre (DIPC), 
POB 1072, 20018 San Sebastian, Spain}
\author{J. von Boehm}
\affiliation{
COMP/Laboratory of Physics, 
Helsinki University of Technology, POB 1100, 02015 HUT, Finland.
}

\date{\today}

\begin{abstract}
The magnetization and hole distribution
of Mn clusters in (Ga,Mn)As are investigated by
all-electron total energy calculations using the projector augmented wave
method within the density-functional formalism. It is found that the
energetically most favorable clusters consist of Mn atoms surrounding one
center As atom. As the Mn cluster grows the hole band at the Fermi
level splits increasingly and the hole distribution
gets increasingly localized at the center As atom. The hole distribution
at large distances from the cluster does not depend significantly
on the cluster size. As a consequence, the spin-flip energy differences
of distant clusters are essentially independent of the cluster size.
The Curie temperature $T_C$ is estimated directly from these spin-flip energies
in the mean field approximation. When clusters are present estimated
$T_C$ values are around 250 K independent of Mn concentration
whereas for a uniform Mn distribution we estimate a $T_C$ of about 600 K.
\end{abstract}
\noindent {\it PACS \# \ }

\pacs{75.50.Pp; 85.75.-d}
\maketitle

\section{Introduction}

In the diluted magnetic semiconductor (Ga,Mn)As the Mn atoms
substituting Ga ones act as acceptors that simultaneously provide both
local magnetic moments and spin-polarized \textit{p}-\textit{d}-type
delocalized holes (according to a simple model one hole per Mn atom) which are
generally believed to mediate the ferromagnetic
coupling~\cite{ohno-ea-1996,matsukura-ea-1998,ohno-1998,ohno-1999,szczytko-ea-1999,beschoten-ea-1999,dietl-ea-2000,ohno-matsukura-2001,akinaga-ohno-2002,kepa-ea-2003,coey-sanvito-2004,dietl-2004}.
This unique feature where magnetism is clearly intertwined with
semiconductor properties makes (Ga,Mn)As a particularly attractive
material both for basic research and spintronics
applications. The Curie temperature ($T_C$) has been found to depend
directly on the delocalized hole concentration~\cite{satoh-ea-1997,ohno-1998,ohno-1999,potashnik-ea-2001,edmonds-ea-2002,yu-ea-2003,ku-ea-2003,sorensen-ea-2003}. It has also been found that annealing of the molecular beam epitaxially grown
(Ga,Mn)As removes Mn interstitial donors which increases delocalized hole
concentration and consequently $T_C$~\cite{yu-ea-2002,yu-ea-2003,edmonds-ea-2004}. The disorder of the Mn atoms either increases $T_C$ further
for lower Mn concentrations~\cite{berciu-bhatt-2001,chudnovskiy-pfannkuche-2002,kaminski-dassarma-2002,sandratskii-ea-2004}
or decreases $T_C$~\cite{xu-ea-2005}.
Another mechanism which affects hole concentration (decreasingly)
is the donation of electrons by the As antisite defects in the as-grown
(Ga,Mn)As~\cite{potashnik-ea-2001,grandidier-ea-2000,sanvito-hill-2001,korzhavyi-ea-2002,kudrnovsky-ea-2004,tuomisto-ea-2004}.

However, an additional mechanism which surely affects magnetic coupling and
the related hole distribution as well as $T_C$ is the clustering of Mn
atoms. Mn$_{\rm Ga}$-Mn$_{\rm Ga}$ dimer clusters (denoted henceforth by Mn$_2$ and
called ``dimer'') and Mn$_{\rm Ga}$ - interstitial Mn complexes have been found
in as-grown (Ga,Mn)As samples from cross-sectional scanning
tunneling microscopy images~\cite{sullivan-ea-2003}. 
Some of these Mn$_{\rm Ga}$ - interstitial Mn complexes may disappear during
annealing~\cite{edmonds-ea-2004}, but complexes where the interstitial Mn
is bound to substitutional Mn clusters can be very stable~\cite{mahadevan-zunger-2003}.
The presence of
Mn$_2$-clusters is natural because for a characteristic Mn concentration
of $x=$ 5 \% and assuming a random Mn distribution the probability that
a Mn atom belongs to a cluster is $1 - 0.95^{12} \approx 0.46$. 
Structural evolution during post-growth annealing can lead to
further clustering~\cite{hayashi-ea-2001,potashnik-ea-2001,yu-ea-2002}
but no precipitation has been found~\cite{hayashi-ea-2001}.
Recently several {\it first-principles} calculations based on the
density-functional-theory have been performed
to study the Mn clustering and the related changes in
magnetism~\cite{schilfgaarde-mryasov-2001,rao-jena-2002,mahadevan-zunger-2003,raebiger-ea-2004,sandratskii-bruno-2004,raebiger-ea-2005,xu-ea-2005}.
The general outcome of these calculations may be summarized as follows.
Substitutional Mn clustering in (Ga,Mn)As is energetically favorable up to
Mn$_3$- or Mn$_4$-clusters~\cite{schilfgaarde-mryasov-2001,raebiger-ea-2005}.
Large magnetic moments are formed at the Mn clusters in
(Ga,Mn)As~\cite{raebiger-ea-2004,sandratskii-bruno-2004}
[see also Ref.~\cite{rao-jena-2002} for the case (Ga,Mn)N)].
At the same time the \textit{p}-\textit{d} hole distribution grows
significantly at the Mn clusters reducing the relative amount of delocalized
holes available for the long range
coupling of the magnetic moments of the Mn
clusters~\cite{raebiger-ea-2004,raebiger-ea-2005}.
Interstitial Mn is also found to form a stable complex
with Mn$_2$~\cite{mahadevan-zunger-2003}.

The aim of this Paper is to study theoretically the electronic and magnetic
properties of substitutional Mn clusters
in (Ga,Mn)As paying special attention to the hole mediated magnetic
coupling between the magnetic moments of the Mn clusters
using {\it first-principles} methods. 
We limit ourselves solely to substitutional Mn clusters;
interstitial Mn, As antisite or other possible point defects 
are not considered although they may influence the physics 
of (Ga,Mn)As~\cite{potashnik-ea-2001,grandidier-ea-2000,sanvito-hill-2001,korzhavyi-ea-2002,kudrnovsky-ea-2004,tuomisto-ea-2004,mahadevan-zunger-2003}.
The Paper is organized as follows.
The computational methods are presented in Sec. II, the results for single
substitutional Mn impurity, Mn cluster formation and magnetic coupling
including discussion are presented in Sec. III, and the conclusions are
drawn in Sec. IV.

\section{Computational methods}

   Spin-polarized total energy supercell calculations based on the
density-functional-theory are performed for (Ga,Mn)As.
The projector augmented-wave method together with the
generalized gradient approximation (GGA-PW91) for exchange-correlation
as implemented in the VASP code is
employed~\cite{perdew-wang-1992,kresse-furthmuller-1996,kresse-joubert-1999}. 
The projector
augmented wave method has the advantage that it is basically an all-electron
method but nevertheless almost as fast as the usual plane-wave pseudopotential
method. 
However, a slight limitation of the VASP implementation is the fact that the
core states are kept fixed. The projector augmented-wave
potentials
and the pseudopotentials for plane waves 
provided in the VASP package
were
compared with an all-electron full-potential linearized
augmented-plane-wave calculation
by
performing calculations for MnAs (a crystal having the GaAs structure but
Mn substituting Ga). 
The projector augmented-wave calculation was found to agree
closely with the linearized augmented-plane-wave calculation but to differ both
quantitatively and qualitatively from the plane-wave pseudopotential
calculation~\cite{raebiger-ea-2004}.

   In our calculations
plane-waves up to the 275 eV cut-off value are included;
total energy in several test systems change less than 1 meV
as the cut-off value is increased from 275 to 300 eV or from 275 to 325 eV.
Since
(Ga,Mn)As is a half-metal a dense k-mesh is needed. 
The Brilloin zone integrations are done using the linear tetrahedron method
with Bloechl corrections~\cite{kresse-furthmuller-1996}.
The Monkhorst-Pack
$(0.14$ \AA$^{-1})^3$ mesh including the $\Gamma$-point was found
sufficiently accurate for the Brillouin zone sampling~\cite{raebiger-ea-2004}
and is used in this study.
One or two Mn clusters (comprising of up to 5 Mn atoms substituting
Ga sites) are included in supercells of 64, 96 or 128 atoms,
corresponding to $2\times2\times2$, $2\times2\times3$ and $2\times2\times4$
cubic zinc-blende unit cells, respectively. Thus, the Mn concentration
in our calculations varies in the experimentally relevant range
[Mn] = 1.6 \ldots 7.8 \%.
Various Mn cluster distributions and different spin alignments
(e.g. $\uparrow\uparrow$ vs. $\uparrow\downarrow$) are studied.
Since relaxation effects were found negligible in the 64 atom
supercell calculations, only unrelaxed lattice positions with the fixed
experimental lattice constant $a =$ 5.65 {\AA} are used for larger supercells.
The orbital decomposition analysis is performed using the standard methods
implemented in the VASP code~\cite{kresse-furthmuller-1996}. The
self interaction error and the spin-orbit interactions are expected to
be small for (Ga,Mn)As~\cite{perdew-zunger-1981,filippetti-ea-2004,wierzbowska-ea-2005,dasilva-ea-2004} and thus neglected in our calculations.

\section{Results and discussion}

\subsection{Mn$_{\rm Ga}$ impurity in the dilute limit}

\begin{figure*}[hbtp!]
\begin{center}
\epsfig{file=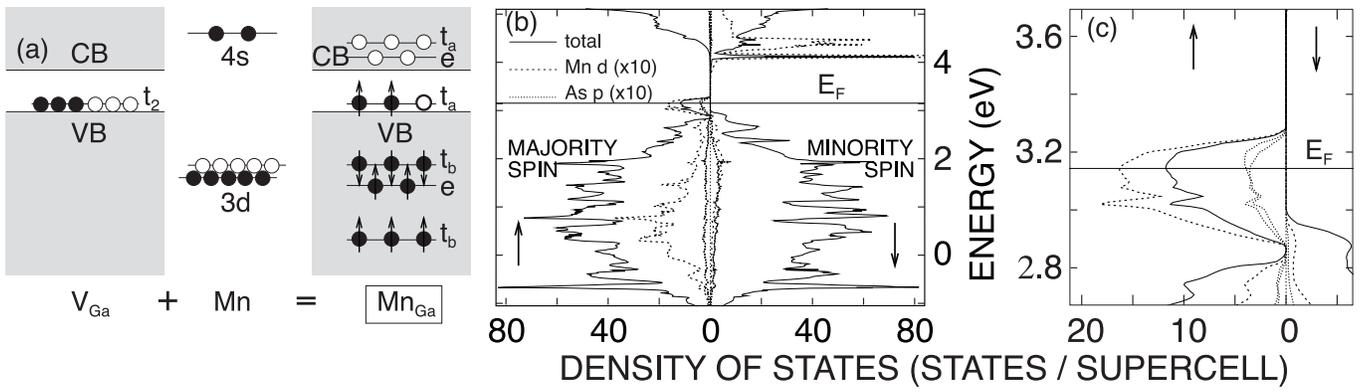,width=\linewidth}
\caption{The calculated density of states (DOS) for the Mn concentration
of $x = 1.6$ \%. (a) Branching diagram for a substitutional Mn atom.
The states of the substitutional Mn atom are formed via the hybridization
of the V$_{\rm Ga}$ $t_2$ and Mn $3d$ states. (b) DOS. $E_F$ denotes the
Fermi level. (c) The magnification of the DOS around $E_F$.
}
\label{fig:single}
\end{center}
\end{figure*}

   The calculated density of states (DOS) for a uniform~\cite{footnote}
substitutional Mn 
distribution in our dilute limit of $x = 1.6$ \% (or one substitutional Mn
atom in the 128 atom supercell) is shown in Fig.~\ref{fig:single} for
reference. The formation of the DOS may be followed using the simple
branching diagram of Fig.~\ref{fig:single} (a). The removal of a
Ga atom creates a vacancy (V$_{\rm Ga}$) acting as a shallow triple acceptor. The
six-fold degenerate \textit{p}-type $t_2$-level of the V$_{\rm Ga}$ is occupied
by three electrons~\cite{laasonen-ea-1992,scherz-scheffler-1992}.
When the Mn atom ($3d^54s^2$) is placed to the V$_{\rm Ga}$, the
V$_{\rm Ga}$ \textit{p} states and the Mn \textit{d} states hybridize
into the $t_{b\uparrow}$ \& $t_{b\downarrow}$ bonding states, the
$e_{\uparrow}$ \& $e_{\downarrow}$ states, and the $t_{a\uparrow}$
\& $t_{a\downarrow}$ antibonding states. The three electrons from the V$_{\rm Ga}$
and the seven electrons from the Mn atom occupy the
three $t_{b\uparrow}$ bonding states, the two $e_{\uparrow}$ states,
the three $t_{b\downarrow}$ bonding states, and two of the
three $t_{a\uparrow}$ antibonding states leaving one unoccupied
state (Fig.~\ref{fig:single} (a)). The bands and the DOS are subsequently
formed as shown in Fig.~\ref{fig:single} (b).
The most important feature of the DOS is the single hump in the majority
spin ($\uparrow$) channel at the Fermi level ($E_F$) which makes (Mn,Ga)As
a half-metal (Figs~\ref{fig:single} (b) and (c)).
As discussed above, the hump is formed mainly from the
$t_{a\uparrow}$ antibonding states (Fig.~\ref{fig:single} (a)).
Each Mn atom contributes one hole state
in the unoccupied part of the hump as well as the net magnetic moment of
4$\mu_B$ [two $e_{\uparrow}$ and two $t_{a\uparrow}$ electrons, see
Fig.~\ref{fig:single} (a)] ($\mu_B$ is the Bohr magneton).
Our calculated DOS in Fig.~\ref{fig:single} agrees closely with several
other independent calculations~\cite{sato-katayama-yoshida-2001,sanvito-ea-2001,kulatov-ea-2002,sato-ea-2003,bergqvist-ea-2003,sandratskii-ea-2004,wierzbowska-ea-2005}.
\begin{figure}[htbp!]
\begin{center}
\psfig{file=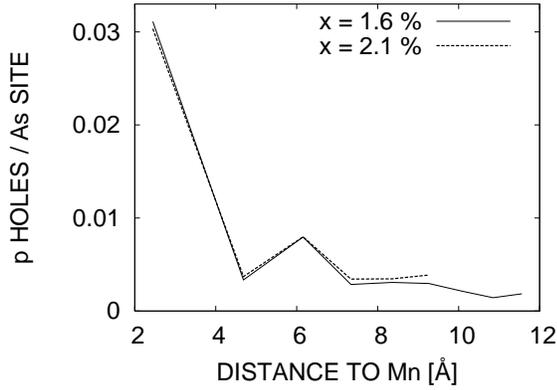,angle=-90,width=0.9\linewidth}
\caption{The calculated As $p$ hole distribution. The average number
of $p$ holes per As site at each As coordination cell is given. The numbers are obtained
by integration from the orbital decomposition. Note that the orbital
decomposition as well as the choice of the integration volume are not
unambiguous. Nevertheless, the relative magnitudes are directly comparable.
}
\label{fig:hole-loc}
\end{center}
\end{figure}

   The spin-polarized hole density of the hump consists mainly of the Mn
$d$ part localized around the Mn atoms and the As $p$ part that is
\textit{delocalized} around the As atoms. This delocalization
can be seen in the calculated As $p$ projection of the hole distribution
shown in Fig.~\ref{fig:hole-loc} (the average number of $p$ holes
at the As atoms belonging to each coordination shell measured from
the closest Mn site is given; for comparison the number
of holes at the Mn site is about 0.13).
The peaks at the distances of 2.4 and 6.2 {\AA} belong
to the nearest and third nearest As coordination shells.
Otherwise the $p$ hole distribution appears to be quite even. 
The $p$ hole distribution does not change significantly when the Mn
concentration is changed (cf. the solid and dotted lines in
Fig.~\ref{fig:hole-loc} corresponding to the Mn concentrations of
1.6 and 2.1 \%, respectively).

\subsection{Formation of Mn clusters}

\begin{figure}[tbp!]
\begin{center}
\epsfig{file=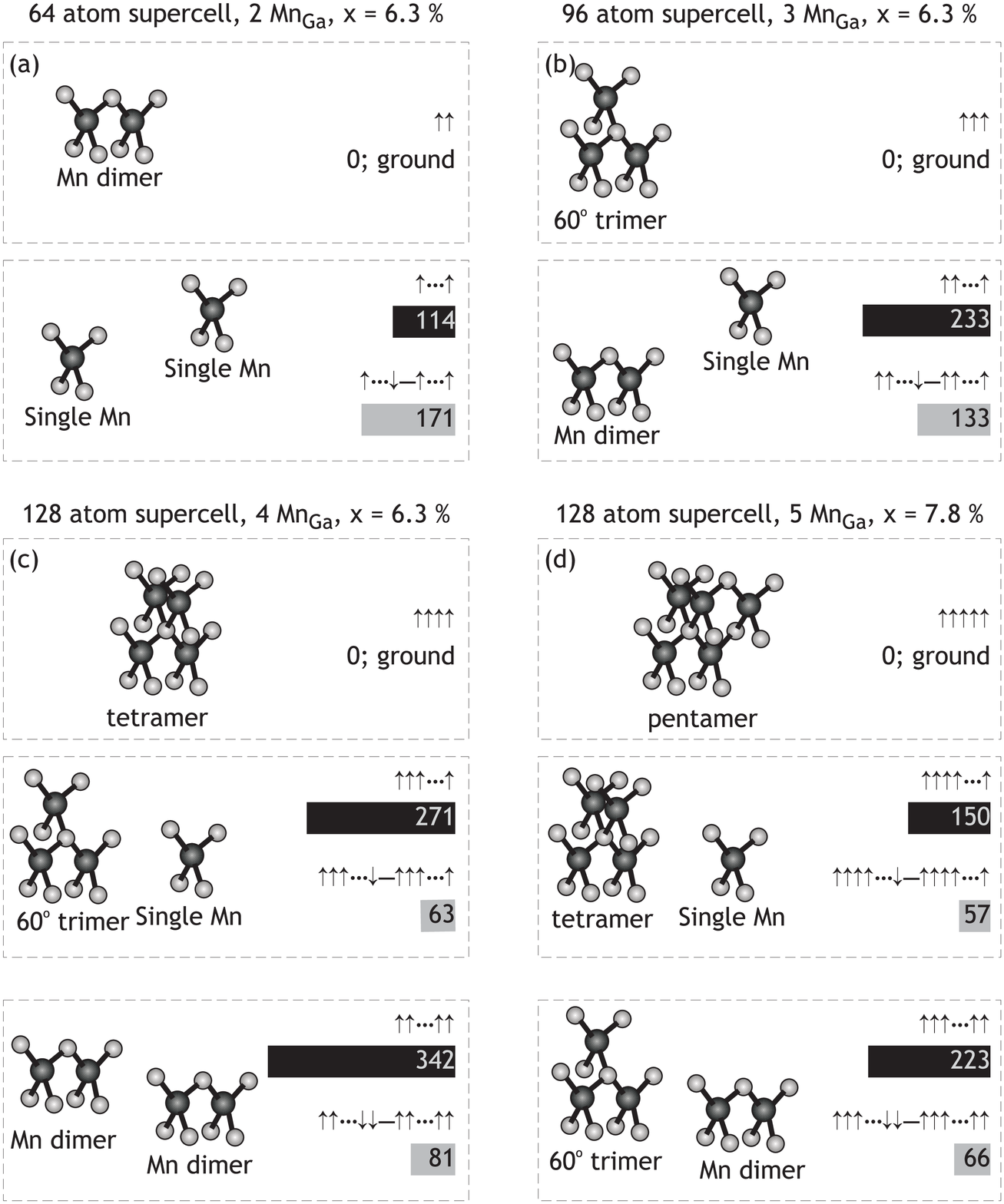,width=\linewidth}
\caption{
Calculated Mn cluster configurations and total energies. The dark and gray
balls denote the Mn and As atoms, respectively. Separated components are placed
at the maximum distance available. All the black horizontal bars show the
(ferromagnetic) separation (or binding) energies; the corresponding values
are given in the units of meV in the black bars. The gray horizontal bars
with the arrow diagrams show the spin-flip energies; the corresponding values
are given in the units of meV inside the gray bars. All the black and gray
horizontal bars are drawn in the same scale.
}
\label{fig:megatable}
\end{center}
\end{figure}

The energetically most important calculated Mn cluster
configurations are shown in Fig.~\ref{fig:megatable}. The supercells used are
chosen such that the Mn concentration remains the same $x =$ 6.3 \%. However,
in the case of five substitutional Mn atoms in the supercell a slightly higher
concentration of 7.8 \% is allowed due to the computational limitations.
For the cases of 2-4 Mn atoms in the supercell the configuration
where the Mn atoms share the \textit{same
neighbouring} As site is always found to be energetically most favorable
(see the upper left-hand corners in Figs~\ref{fig:megatable} (a) - (c)).
Also, it is energetically favorable to form one cluster from two separate
components in all cases shown in Figs~\ref{fig:megatable} (a) - (c).
In the case of five Mn atoms in the supercell the energetically
most favorable configuration is obtained by attaching the fifth Mn atom
to the stable tetramer [Fig.~\ref{fig:megatable} (d)]. 
The energies needed to separate single clusters into two components
(or the binding energies for the components) are shown graphically
in Fig.~\ref{fig:megatable} as black horizontal bars which contain
also the corresponding energy values in the units of meV.
In all these cases the ferromagnetic order is the stablest magnetic phase.
By using the above
binding energies we have calculated the heats of reaction for the cluster
formation and found that the optimal cluster sizes are
tetramers being slightly more favorable than the
trimers~\cite{raebiger-ea-2005}. Our calculation showing that clustering
is energetically favorable is in agreement
with the experimental result that Mn$_2$ clusters appear already in the
as-grown (Ga,Mn)As samples~\cite{sullivan-ea-2003}. Further clustering
is expected during post-growth annealing.

\subsection{Effects of clustering}

\begin{figure*}[hbtp!]
\begin{center}
\epsfig{file=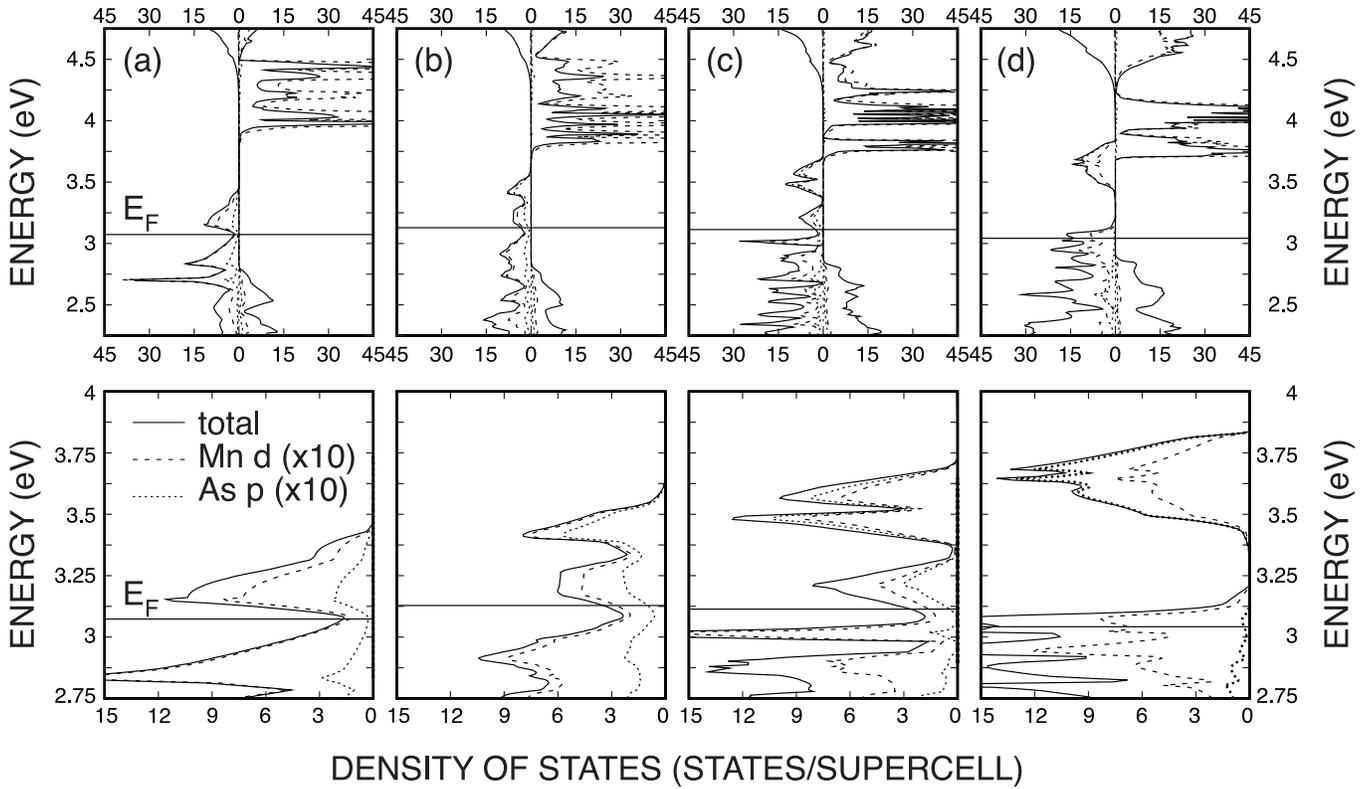,width=\linewidth}
\caption{Densities of states (DOSs) around the band gap (upper figures) and
the magnifications of the majority spin DOSs around the Fermi energy ($E_F$)
(lower figures). Figures (a), (b), (c), and (d) give the DOSs for the
Mn monomer, dimer, trimer, and tetramer, respectively for
the Mn density of $x = 6.3 \%$.
}
\label{fig:dostrends}
\end{center}
\end{figure*}

The dilute-limit majority-spin hump at the Fermi energy $E_F$ in
Fig.~\ref{fig:single} (c) is found to split when Mn clusters
are formed, and new narrower unoccupied bands appear in the gap.
This is shown for the Mn monomer, dimer, trimer, and tetramer systems
at the constant Mn concentration of $x = 6.3 \%$ in Fig.~\ref{fig:dostrends}.
At the same time the hole density grows at
the As atom which is situated in the center of the Mn cluster. This
was shown for the Mn dimer in Ref.~\cite{raebiger-ea-2004}
and a similar splitting was also found in Ref.~\cite{sandratskii-bruno-2004}.
The increasing localization at the center As atom is reflected
in the increasing separation and size of the
split-off part of the hump in Fig.~\ref{fig:dostrends}.
The As $p$-projection in the split-off part of the hole DOS is seen
to increase in relation to the Mn $d$-projection in Fig.~\ref{fig:dostrends}
and even to exceed the Mn $d$ projection in the cases of the Mn trimer and
tetramer [Figs~\ref{fig:dostrends} (c) and (d)].

   The integrated numbers of the Mn $d$ holes at the cluster Mn atoms
and the As $p$ holes at the center As atom of the clusters are given in
Table~\ref{tab:hole-loc}. The results in Table~\ref{tab:hole-loc} clearly
show that the number of the As $p$ holes grows relatively faster than
the number of the Mn $d$ holes as the cluster size increases.
In the case of the pentamer,
the fifth Mn atom lies outside the tetramer and, in addition to
the center As atom inside the tetramer, there are two equivalent
As atoms between the fifth Mn atom and the tetramer
[Fig.~\ref{fig:megatable} (d)].
In this case the holes in the tetramer part remain intact
whereas the numbers of the Mn $d$ and As $p$ holes
are seen to grow at the additional fifth Mn atom and the two associated
As atoms, respectively (see the last line of Table~\ref{tab:hole-loc}).
We expect that an additional sixth Mn atom would cause further hole
growth at the new center-As-atom.     

\begin{figure}[htbp!]
\begin{center}
\psfig{file=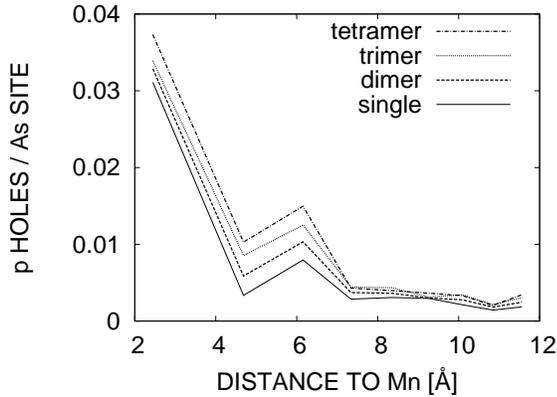,angle=-90,width=0.9\linewidth}
\caption{The calculated As $p$ hole distribution for different cluster sizes.
For further information see the caption of Fig.~\ref{fig:hole-loc}.}
\label{fig:clus-hole-loc}
\end{center}
\end{figure}

  The As $p$ component of the spin-polarized hole density as a function of
the distance to the closest Mn atom is shown in Fig.~\ref{fig:clus-hole-loc}
for several different cluster sizes. At short distances $r < 7$ {\AA}
to the closest Mn atom of the cluster, the hole concentration
increases with increasing number of Mn atoms.
However, for larger distances $r > 7$ {\AA} the hole concentrations
approach the same curve.
Therefore,
the long-distance magnetic coupling between the Mn clusters
depends only on their mutual distance, not on the size of the cluster.

\begin{table}
\caption{Holes inside the Mn cluster. The numbers of the
Mn $d$ holes at the cluster Mn atoms and the As $p$
holes at the As atom which is the center of the Mn
cluster are given. The numbers are obtained
by integration from the orbital decomposition. Although the orbital
decomposition as well as the choice of the integration volume are not
unambiguous the given average numbers of $p$ holes are directly comparable.
}
\label{tab:hole-loc}
\begin{tabular}{l c c c}
\hline
\hline
System & Mn $d$ holes & As $p$ holes & $x$ (\%) \\
\hline
single Mn & 0.13 & 0.03 & 6.3 \\
dimer  & 0.17  & 0.11 & 6.3 \\
trimer  & 0.19  & 0.19 & 6.3 \\
tetramer  & 0.21 & 0.29 & 6.3 \\
pentamer  & 0.21 ($\times$4) & 0.28 ($\times$1) & 7.8 \\
          & 0.19 ($\times$1) & 0.10 ($\times$2) &     \\
\hline
\hline
\end{tabular}
\end{table}

   In addition to the different ferromagnetic cluster configurations above
also various other magnetic alignments were considered. As mentioned above,
the ferromagnetic alignment is always energetically most stable.
The intra-cluster spin-flip energy is relatively high
for the Mn dimer: 211 meV~\cite{raebiger-ea-2004}
and significantly higher inside the larger clusters, typically 300 meV and
more. The spin flip energy of a distant component describes the stability
of the long range ferromagnetic state and may also be used
estimating $T_C$ as will be discussed shortly.
The calculated spin-flip energies $\Delta$ for the spin-flips of distant Mn
atoms or entire clusters are given in Table~\ref{tab:1} as well as in
Fig.~\ref{fig:megatable} as gray horizontal bars containing also the
corresponding energy values in the units of meV. The Mn concentrations
$x$ in Table~\ref{tab:1} are chosen such that the closest Mn-Mn distance
is constant (10.6 {\AA}) except for the first row where the distance is
13.8 {\AA}. The calculated $\Delta$
values vary relatively little, between 57 and 81 meV. 
This variation in the $\Delta$ values exhibits no evident trend;
e.g. by looking at configurations with one cluster fixed, 
an increase of the second cluster size leads to either increase or decrease
of the $\Delta$ value.
These fluctuations may be caused by e.g. directional 
effects~\cite{sanyal-ea-2003,mahadevan-ea-2004}.
This shows clearly
that new Mn atoms in the clusters promote the holes mainly to the center
As atom of the cluster and thus do not increase the delocalized hole
concentration mediating the ferromagnetic cluster-cluster coupling.

 The following expression based on the mean-field approximation
may be used to roughly estimate
$T_C$~\cite{sato-ea-2003,kurz-ea-2002,turek-ea-2003}
\begin{equation}
T_C = \frac{2}{3 k_B}\frac{\Delta}{N},
\label{eq:tc}
\end{equation}
where $\Delta$ is the energy difference between the spinglass and
ferromagnetic arrangements in the supercell, and $N$ denotes the number
of the magnetic particles in the supercell. However, we choose to
calculate $\Delta$ as the difference between the anti-parallel
and parallel arrangements which is known to be a good approximation
~\cite{kurz-ea-2002,turek-ea-2003}.
We treat the Mn clusters as single magnetic
particles, and calculate $\Delta$ from the cluster-cluster
spin-flip energies. The number of Mn clusters
in the supercell is alwaýs fixed to $N = 2$.

   First, we consider the simple antiferromagnetic-ferromagnetic
case. Using Eq.(~\ref{eq:tc}) we estimate that for a uniform Mn
monomer distribution $T_C$ is 220 K for $x = 3.1 \%$
(Table~\ref{tab:1}) and 620 K for $x = 6.3 \%$~\cite{raebiger-ea-2004}.
The increase of $T_C$ from 220 to 620 K is mainly due to the decrease of the
Mn-Mn distance from 13.8 to 9.8 {\AA}. 
If at $x=6.3 \%$ the Mn atoms were distributed randomly
at the Ga sites, about $1-0.937^{12} = 54 \%$ of the Mn atoms
would belong to clusters.
Therefore it is of some interest to compare the uniform Mn-dimer
distribution at $x=6.3\%$ (all Mn atoms belong to dimers)
with the uniform Mn monomer distribution at $x=6.3 \%$.
The dimerization reduces $T_C$ from 620 K to 313 K (Table~\ref{tab:1}).
A similar dramatic decrease of $T_C$ is found at $x=12.5\%$ from a uniform
Mn distribution to the corresponding Mn$_4$ cluster distribution
in Ref.~\cite{sandratskii-bruno-2004},
but at $x=6.3\%$ a slight increase in $T_C$ is found from
a uniform Mn distribution to the corresponding dimer distribution.
In Ref.~\cite{xu-ea-2005} it was concluded quite generally that
clustering decreases $T_C$, which is in agreement with our findings.

   We follow the same procedure as above to estimate $T_C$ from
Eq.(~\ref{eq:tc}) for further cluster distributions. The estimated $T_C$
values are given in Table~\ref{tab:1} where the Mn concentrations $x$ are
chosen such that the closest Mn-Mn distance is always 10.6 {\AA} (except
for the Mn$_1$+Mn$_1$ case). As noted above $\Delta$ does not depend
significantly on the Mn concentration and therefore neither does $T_C$:
the calculated $T_C$ values vary between 220 and 313 K (Table~\ref{tab:1}).
This may be related to the fact that the average As $p$ hole distribution at large
distances ($> 7$ \AA) is almost independent of the 
number of Mn atoms in each cluster. Thus, as long as the
distance between clusters is kept constant, $T_C$ does not vary
significantly though the number of Mn atoms in the clusters is changed.

   In the mean field approximation, $T_C$ is proportional to $\Delta/N$ 
(Eq.(~\ref{eq:tc})). During clustering, $\Delta$ may be expected to decrease
exponentially with respect to the growing cluster-cluster distance
while the number of clusters $N$ in the denominator decreases
much more slowly. Thus, as already noted above for the uniform dimer
distribution, clustering decreases $T_C$ significantly.

   With the present high quality (Ga,Mn)As samples where the amount of
the harmful interstitial Mn and As$_{\rm Ga}$ defects has been minimized
one achieves $T_C$'s in the range
of 159-173 K~\cite{edmonds-ea-2004,wang-ea-2004}. The mean
field approximation used here may be expected to give an upper limit in
estimating $T_C$. The obtained estimates ranging between 220 and 313 K in
the case of clustering (Table~\ref{tab:1}) agree quite reasonably with
these best experimental values. 
This is consistent with the fact that
real samples contain clusters~\cite{sullivan-ea-2003}
which, as was shown above, can
reduce $T_C$ strongly.


\begin{table}
\caption{Spin-flip total energy differences
between two substitutional Mn clusters.
$\Delta$ is the energy
difference between the system with one of the two clusters in anti-parallel
spin state (A) with respect to the system with all spins ferromagnetically
aligned (FM). The 128 atom supercell is used.
}
\label{tab:1}
\begin{tabular}{l c c c c c c}
\hline
\hline
System & FM & & A & $\Delta$ (meV) & $T_C$ (K) & $x$ (\%) \\
\hline
\hline
Mn$_1$+Mn$_1$ & $\uparrow\cdot\cdot\uparrow$ & & $\uparrow\cdot\cdot\downarrow$ & 57 & 220 &3.1 \\
\hline
Mn$_2$+Mn$_1$ & $\uparrow\uparrow\cdot\cdot\uparrow$ & & $\uparrow\uparrow\cdot\cdot\downarrow$ & 73 & 282 & 4.7 \\
\hline
Mn$_3$+Mn$_1$ & $\uparrow\uparrow\uparrow\cdot\cdot\uparrow$ & & $\uparrow\uparrow\uparrow\cdot\cdot\downarrow$ & 63 & 244 & 6.3 \\
Mn$_2$+Mn$_2$ & $\uparrow\uparrow\cdot\cdot\uparrow\uparrow$ & & $\uparrow\uparrow\cdot\cdot\downarrow\downarrow$ & 81 & 313 & 6.3 \\
\hline
Mn$_4$+Mn$_1$ & $\uparrow\uparrow\uparrow\uparrow\cdot\cdot\uparrow$ & & $\uparrow\uparrow\uparrow\uparrow\cdot\cdot\downarrow$ & 57 & 220 & 7.8 \\
Mn$_3$+Mn$_2$ & $\uparrow\uparrow\uparrow\cdot\cdot\uparrow\uparrow$ & & $\uparrow\uparrow\uparrow\cdot\cdot\downarrow\downarrow$ & 66 & 255 & 7.8 \\
\hline
\hline
\end{tabular}
\end{table}

\section{Conclusion}

Substitutional Mn clustering in the diluted (Ga,Mn)As magnetic
semiconductor is studied by means of spin-polarized all-electron
density-functional calculations. The clustering is an important
factor in the typical Mn concentration range from 1 to 10 \% because
already in the case of purely random Mn distribution the probability
that a Mn atom belongs to a cluster varies correspondingly from 0.1 to 0.7. 
Furthermore, our calculations show that cluster formation is energetically
favorable. The energetically most stable clusters are found to consist of
Mn atoms that surround symmetrically the center As atom. The spin-polarized
hole distribution is found to get increasingly localized at this
center As atom and the majority spin DOS hump 
at the Fermi level to split
when the cluster size increases from Mn$_1$ to Mn$_4$.
At the same time the hole density outside the cluster 
depends relatively little on the number of Mn atoms in the cluster,
especially at larger distances ($> 1.2 \times$ lattice constant).
This implies that the long range ferromagnetic coupling
between two Mn clusters depends relatively little on the number
of Mn atoms in the clusters. Our calculated spin-flip energies
confirm this expectation. Using these spin-flip energies and the mean
field approximation we estimate Curie temperatures around 250 K for the
various studied cluster distributions independently of the Mn concentration.
In contrast, we estimate for a uniform Mn monomer distribution at
$x =$ 6.3 \% a high value of 620 K. The best present (Ga,Mn)As
samples achieve Curie temperatures of 159-173 K. Therefore, our
estimated Curie temperature values are consistent with the fact that the (Ga,Mn)As
samples are hampered by the clustering effect.

\section{Acknowledgments}

This work has been supported by the Academy of Finland through the Center of
Excellence Program (2000-2005). A.A. acknowledges the financing of the Basque 
Government by the ETORTEK program called NANOMAT. 
The authors thank Acad. Prof. R. M. Nieminen,
Prof. K. Saarinen, Prof. M. J. Puska, Dr. M. Alava, Mr. F. Tuomisto and Mr.
T. Hynninen for many valuable discussions. We acknowledge the generous
computing resourches of the Center for Scientific Computing (CSC).



\begin{references}
\bibitem{footnote}
In this Paper the uniform Mn distribution means a regular,
periodic Mn sublattice with no clustering, calculated within the supercell approach.
This does not correspond to the homogeneous disorder
in the coherent potential approximation calculations.
\end{references}
\end{document}